\documentclass[prd,twocolumn,superscriptaddress,altaffilletter,amssymb,showpacs,nofootinbib]{revtex4}
\usepackage[dvips]{graphicx}
\usepackage{amsmath}
\newcommand{\be}{\begin{equation}}
\newcommand{\ee}{\end{equation}}
\newcommand{\bea}{\begin{eqnarray}}
\newcommand{\eea}{\end{eqnarray}}

\begin{document}

\title{Critical comments on the paper "Crossing $\omega=-1$ by a single scalar field on a Dvali-Gabadadze-Porrati brane" by H Zhang and Z-H Zhu [Phys.Rev.D75,023510(2007)]}

\author{Israel Quiros}\email{israel@uclv.edu.cu}
\affiliation{Departamento de F\'{\i}sica, Universidad Central de
Las Villas, 54830 Santa Clara, Cuba.}
\author{Ricardo Garc\'{\i}a-Salcedo}\email{rigarcias@ipn.mx}
\affiliation{Centro de Investigacion en Ciencia Aplicada y Tecnologia Avanzada - Legaria del IPN, M\'exico D.F., M\'exico.} \author{Claudia Moreno}\email{claudia.moreno@cucei.udg.mx}
\affiliation{Departamento de F\'{\i}sica y Matem\'aticas, Centro Universitario de Ciencias Ex\'actas e Ingenier\'{\i}as, Corregidora 500 S.R., Universidad de Guadalajara, 44420
Guadalajara, Jalisco, M\'exico}

\date{\today}
\begin{abstract}
It is demonstrated that the claim in the paper "Crossing $\omega=-1$ by a single scalar field on a Dvali-Gabadadze-Porrati brane" by H Zhang and Z-H Zhu [Phys.Rev.D75,023510(2007)], about a prove that there do not exist scaling solutions in a universe with dust in a Dvali-Gabadadze-Porrati (DGP) braneworld scenario, is incorrect.
\end{abstract}

 
\maketitle

Since the discovery that our universe can be currently undergoing a stage of accelerated expansion \cite{obs}, many phenomenological models based either on Einstein General Relativity (EGR), or using alternatives like the higher dimensional brane world theories \cite{roy}, have been invoked (for a recent review on the subject see reference \cite{Copeland:2006wr}). The latter ones, being phenomenological in nature, are inspired by string theory.

One of the brane models that have received most attention in recent years is the so called Dvali-Gabadadze-Porrati (DGP) brane world \cite{dgp}.\footnotemark\footnotetext{For cosmology of DGP braneworlds see reference \cite{dgpcosmology}.} This model describes a brane with 4D world-volume, that is embedded into a flat 5D bulk, and allows for infrared/large scale modifications of gravitational laws. A distinctive ingredient of the model is the induced Einstein-Hilbert action on the brane, that is responsible for the recovery of 4D Einstein gravity at moderate scales, even if the mechanism of this recovery is rather non-trivial \cite{deffayet}. The acceleration of the expansion at late times is explained here as a consequence of the leakage of gravity into the bulk at large (cosmological) scales, so it is just a 5D geometrical effect,unrelated to any kind of misterious "dark energy". 

The study of the dynamics of DGP models is a very atractive subject of research. It is due, in part, to the very simple geometrical explanation to the "dark energy problem", and, in part, to the fact that it is one of a very few possible consistent infrared modifications of gravity that might be ever found. In particular, there can be found studies of the dynamics of a self-interacting scalar field trapped on a DGP brane by invoking the dynamical systems tools, which have been proved useful to retrieve significant information about the evolution of a huge class of cosmological models. In this regard, the exponential potential represents a common functional form for self-interaction potentials that can be found in higher-order \cite{witt} or higher-dimensional theories \cite{nd}. These can also arise due to non-perturbative effects \cite{nonp}. 

A dynamical study of DGP models with a self-interacting scalar field trapped on the DGP brane has been undertaken, for instance, in reference \cite{zhang} for an exponential potential, to show that crossing of the phantom barrier $\omega=-1$ is indeed possible in DGP cosmology with a single scalar field (see also \cite{clmq} in this regard). However, the authors of that paper do not study in detail the phase space of the model and, in correspondence, they are not able to find critical points. Their claim that scaling solutions do not exist in a universe with dust on a DGP brane (only the Minkowski cosmological phase is considered), seems to be in contradiction with known results. In fact, in the 4D limit (the formal limit when, in the DGP model, the crossover length $r_c=k_5^2\mu^2\rightarrow\infty$)\footnote{We follow here the same simbology and terminology used in reference \cite{zhang}.} the results of reference \cite{wands} have to be recovered, or, at least, approached, since the investigation in \cite{zhang} is just a generalization of the one reported in \cite{wands}, to include higher-dimensional behaviour dictated by the DGP dynamics. Even if one expects that the late-time structure of the phase space is modified by the contribution of the DGP brane, these modifications should be associated with the stability of the corresponding critical points rather than with their mere existence. 

In the present comment we show which is the source of the incorrect result of reference \cite{zhang}, and we perform an exhaustive analysis of the phase space for the DGP model with a self-interacting scalar field trapped on the brane, by using the same variables of \cite{zhang}. It is shown, in particular, that there is actually an isolated critical point associated with the matter-scaling solution, even if it is always a saddle point in phase space.

\begin{table*}[tbp]\caption[crit1]{Critical points of the autonomous system of differential equations (\ref{22}-\ref{25}).}
\begin{tabular}{@{\hspace{4pt}}c@{\hspace{14pt}}c@{\hspace{14pt}}c@{\hspace{14pt}}c@{\hspace{14pt}}c@{\hspace{14pt}}c@{\hspace{14pt}}c@{\hspace{14pt}}c}
\hline\hline\\[-0.3cm]
$P_i$ &$x$&$y$&$b$&Existence&
 $\omega_\phi$& $q$ & $l^2$\\[0.1cm]\hline\\[-0.2cm]
$P_1$& $0$&$0$&$0$ & All $\lambda$ &undefined&$1/2$&$1$\\[0.2cm]
$P_2^\pm$& $\pm 1$&$0$&$0$ & All $\lambda$&$1$& $2$&$0$\\[0.2cm]
$P_3$&$\sqrt\frac{3}{2}\frac{1}{\lambda}$&$\sqrt\frac{3}{2}\frac{1}{\lambda}$&$0$ & $\lambda^2>3$ &$0$& $1/2$&$1-\frac{3}{\lambda^2}$\\[0.2cm]
$P_4$& $\frac{\lambda}{\sqrt 6}$&$\sqrt{1-\frac{\lambda^2}{6}}$&$0$ & $\lambda^2<6$&$\frac{\lambda^2}{3}-1$& $\frac{\lambda^2}{2}-1$&$0$\\[0.4cm]\hline \hline
\end{tabular}\label{tab1}
\end{table*}

The starting point is the Friedmann-DGP equation on the brane (equation (5) of reference \cite{zhang}):

\be H^2+\frac{k}{a^2}=\frac{1}{3\mu^2}\left[\rho+\rho_0+\theta\rho_0\left(1+\frac{2\rho}{\rho_0}\right)^{1/2}\right],\label{5}\ee where $H=\dot a/a$ is the Hubble parameter, $a$ is the scale factor, $k$ is the spatial curvature of the three-dimensional (maximally symmetric) Friedamnn-Robertson-Walker (FRW) space -- taken here to be vanishing: $k=0$ -- , and $\theta=\pm 1$ denotes the two branches of the DGP model (the two possible ways to embedd the DGP brane into the Minkowski bulk). In what follows we shall consider the Minkowski cosmological phase, i. e., the case $\theta=-1$, exclusively. The total energy density on the brane $\rho$, includes dust matter and the scalar field "fluid":

\be \rho=\rho_\phi+\rho_m.\label{6}\ee The effective "density" $\rho_0$ relates the strength of 5-dimensional gravity with respect to the 4-dimensional gravity,

\be \rho_0=\frac{6\mu^2}{r_c^2},\label{7}\ee where, as usual, $r_c$ is the crossover radius. It is evident that 4-dimensional behavior is associated with the formal limit $\rho_0\rightarrow 0$. In the model of interest,

\be \rho_\phi=\frac{1}{2}\dot\phi^2+V(\phi),\;\;p_\phi=\frac{1}{2}\dot\phi^2-V(\phi),\label{1415}\ee where a dot denotes derivative with respect to the cosmic time $t$ and $V(\phi)$ is the self-interaction potential, taken here to be in the form of a single exponential:

\be V(\phi)=V_0\;e^{-\lambda(\phi/\mu)}.\label{16}\ee Here $\lambda$ is a constant parameter and $V_0$ denotes the initial value of the potential.

In order to write the equations of the present model: the Friedmann equation (\ref{5}) plus the continuity equations

\be \dot\rho_m+3H\rho_m=0,\;\;\dot\rho_\phi+3H(\rho_\phi+p_\phi)=0,\label{continuity}\ee in the form of a (autonomous) dynamical system, the following dynamical variables are chosen (see equations (18)-(21) in \cite{zhang}):

\be x\equiv\frac{\dot\phi}{\sqrt 6\mu H},\;y\equiv\frac{\sqrt V}{\sqrt 3\mu H},\;l\equiv\frac{\sqrt{\rho_m}}{\sqrt 3\mu H},\;b\equiv\frac{\sqrt{\rho_0}}{\sqrt 3\mu H}.\label{1821}\ee These variables are subject to the Friedmann constraint (equation (27) in \cite{zhang}) coming from the Friedamnn-DGP equation (\ref{5}):

\be x^2+y^2+l^2+b^2-b^2\left(1+2\frac{x^2+y^2+l^2}{b^2}\right)^{1/2}=1,\label{27}\ee or, alternartively

\be l^2=1-x^2-y^2\pm\sqrt{2}b.\label{constraint}\ee In what follows we shall consider only the "$+$" sign in (\ref{constraint}), since we are focused here on expanding universes only, while the oposite sign "$-$" corresponds to contracting universes. Thanks to the constraint (\ref{constraint}), the dimension of the phase space can be reduced from 4 to 3. The consequence is that, only 3 of the 4 ordinary differential equations (22-25) in \cite{zhang}, are independent. Here we choose the following as independent equations:

\bea &&x'=-\frac{3}{2}\alpha x (2x^2+l^2)+3x-\frac{\sqrt 6}{2}\lambda y^2,\label{22}\\
&&y'=-\frac{3}{2}\alpha y (2x^2+l^2)+\frac{\sqrt 6}{2}\lambda x y,\label{23}\\
&&b'=-\frac{3}{2}\alpha b (2x^2+l^2),\label{25}\eea where

\be \alpha\equiv 1-\left(1+2\frac{x^2+y^2+l^2}{b^2}\right)^{-1/2},\label{26}\ee and the prime denotes derivative with respect to the time variable $s\equiv\ln a$. As already said, due to the constraint (\ref{27}), the variable $l$ can be written as a function of the remaining variables $x$, $y$, and $b$ (see equation (\ref{constraint})), so that the differential equations (\ref{22}-\ref{25}) can be written in the alternative form:

\bea &&x'=-\frac{3}{2}\alpha x (1+x^2-y^2+\sqrt{2}b)+3x-\frac{\sqrt 6}{2}\lambda y^2,\label{22'}\\
&&y'=-\frac{3}{2}\alpha y (1+x^2-y^2+\sqrt{2}b)+\frac{\sqrt 6}{2}\lambda x y,\label{23'}\\
&&b'=-\frac{3}{2}\alpha b (1+x^2-y^2+\sqrt{2}b),\label{25'}\eea where, now

\be \alpha=1-\frac{b}{\sqrt{b^2+2\sqrt{2}b+2}}.\ee

The first step is to identify the phase space for our model, which is given by the non-compact 3-dimensional region:\footnote{Our focus will be on expanding universes only.}

\be \Psi=\{(x,y,b): 0\leq x^2+y^2\leq 1+\sqrt{2}b,0\leq y,0\leq b\},\label{phasespace}\ee 

Note that, a straightforward analysis of the ordinary differential equations in the above autonomous system of equations (\ref{22}-\ref{25}), shows that we can not have $x=y=b=l=0$ simultaneously, since this would imply that the constraint (\ref{constraint}) is not obeyed. Therefore, the critical point found in the reference \cite{zhang}, where $x=y=b=l=0$ at the same time, does not really belong in the phase space $\Psi$ of the model under study. By the same reason, the other point found in \cite{zhang}: $x=y=l=0,b=const$, does not belong in $\Psi$ neither. In fact, from the constraint (\ref{constraint}) it follows that, if $x=y=l=0$, then $b=-1/\sqrt 2$, which is not in $\Psi$ since we are considering expanding FRW universes only.

Going a step forward we can realize that the only critical points of the autonomous system (\ref{22'}-\ref{25'}) are associated with the four-dimensional limit $r_c\rightarrow\infty\;\Rightarrow\;\rho_0\rightarrow 0\;\Rightarrow\;b=0$.\footnote{The critical points $(x,y,b)=(0,0,-1/\sqrt 2)$ and $(0,0,-0.91\pm i\; 0.68)$ do not belong in the phase space $\Psi$ so that these are not critical points of the present model.} This case coincides with the one studied in reference \cite{wands}. The constraint (\ref{constraint}) translates now into the following relationship:

\be l^2=1-x^2-y^2,\;\;\Rightarrow\;\alpha=1.\label{relationship}\ee We are led with the two-dimensional system of equations:

\bea &&x'=-\frac{3}{2}x(1+x^2-y^2)+3x-\frac{\sqrt 6}{2}\lambda y^2,\nonumber\\
&&y'=-\frac{3}{2}y(1+x^2-y^2)+\frac{\sqrt 6}{2}\lambda xy.\label{edo}\eea

\begin{figure}[ht!]
\begin{centering}
\includegraphics[width=4.0cm,height=3.5cm]{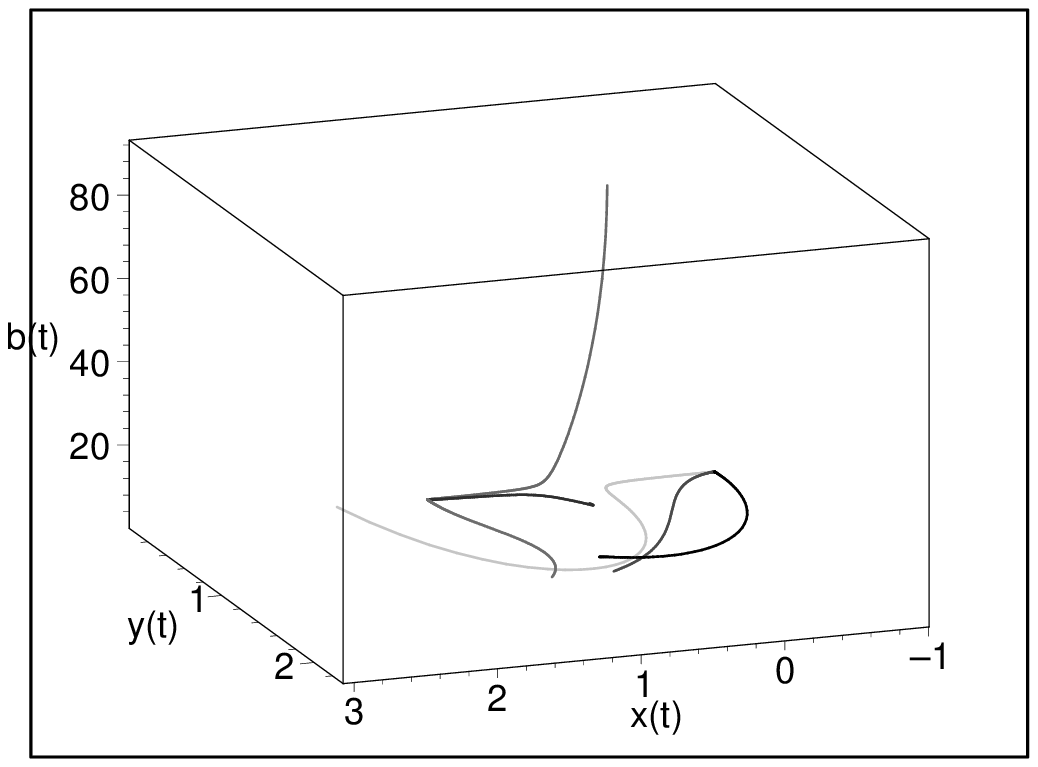}
\includegraphics[width=4.0cm,height=3.5cm]{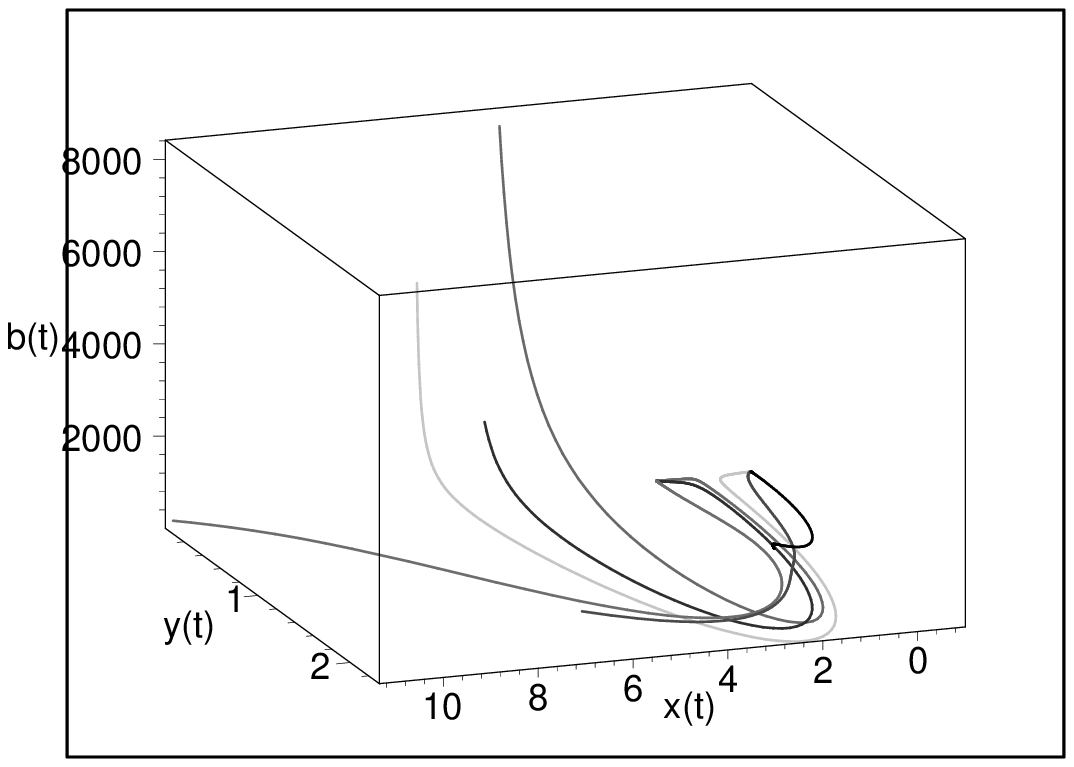}
\includegraphics[width=4.0cm,height=3.5cm]{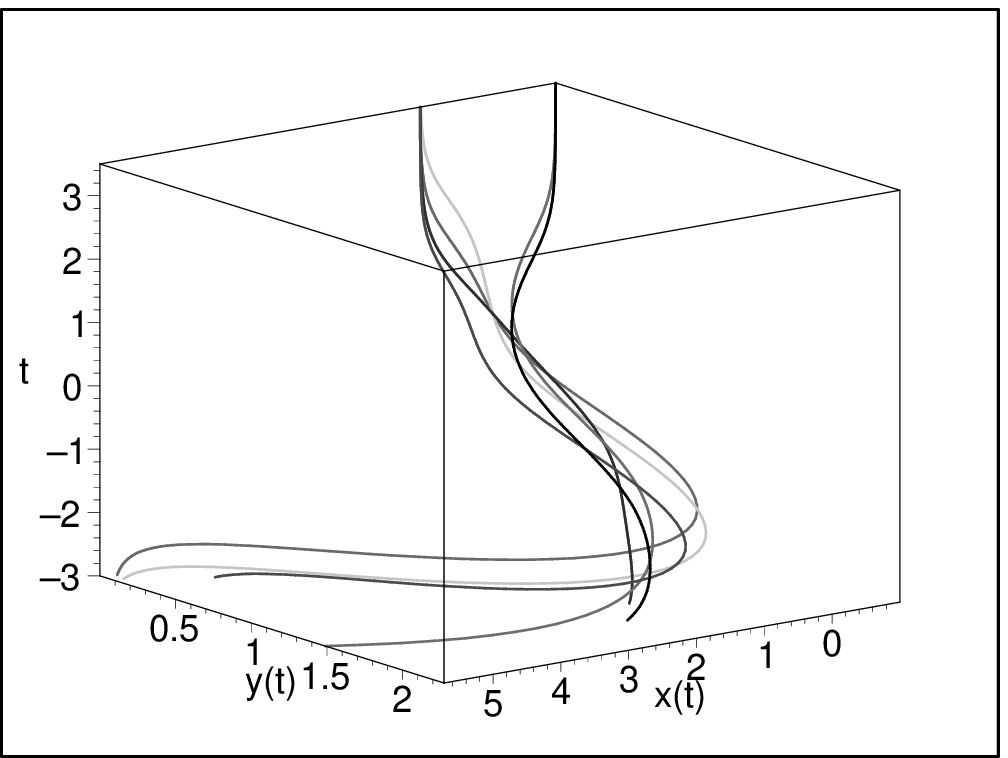}
\includegraphics[width=4.0cm,height=3.5cm]{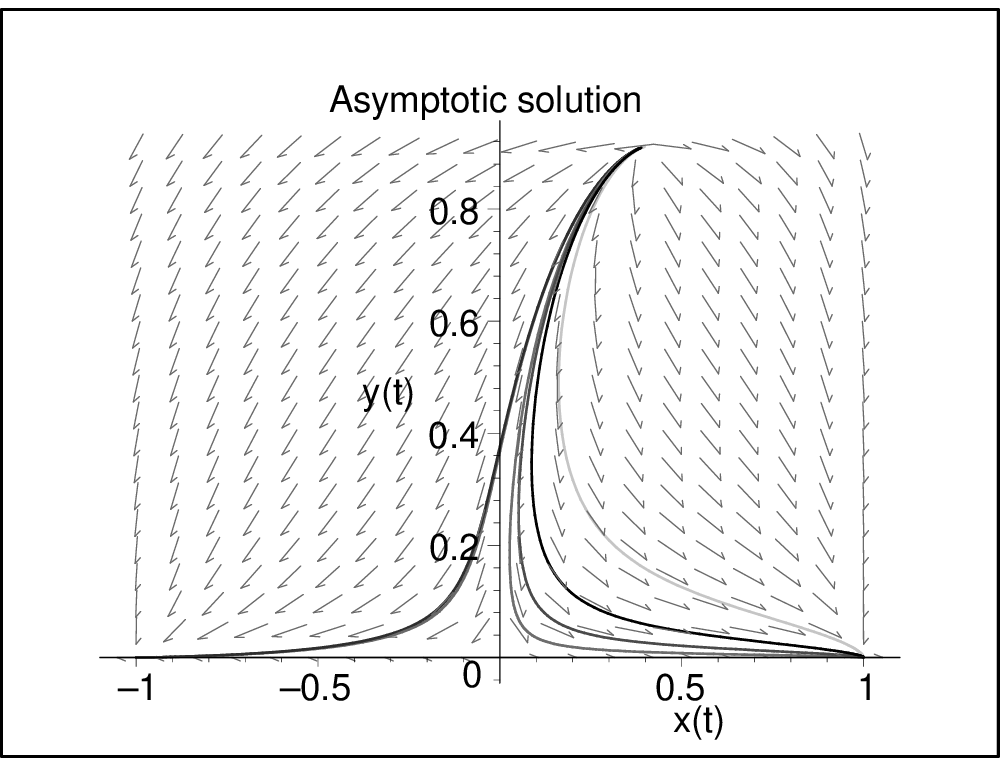}
\vspace{0.3cm}
\caption{Trajectories in phase space for different sets of initial conditions. Pictures in the upper part of the figure show different levels of magnification. The flux in time $\tau$ is shown in the left-hand picture in the lower part of the figure. The value of the parameter of the exponent in the potential $V=V_0 e^{\lambda\phi/\mu}$, has been chosen to be: $\lambda=2$. In the right-hand picture in the lower part of the figure we show the projection onto the plane $(x,y,0)$. The kinetic energy dominated solution $P_2^\pm$ is the past attractor. The scalar field-dominated solution seems to be the future (late-time) attractor. This is just a "mirage" due to the fact that the brane effects are ignored in the projection.}\label{fig1}
\end{centering}
\end{figure}

\begin{figure}[t!]
\begin{centering}
\includegraphics[width=4.0cm,height=3.5cm]{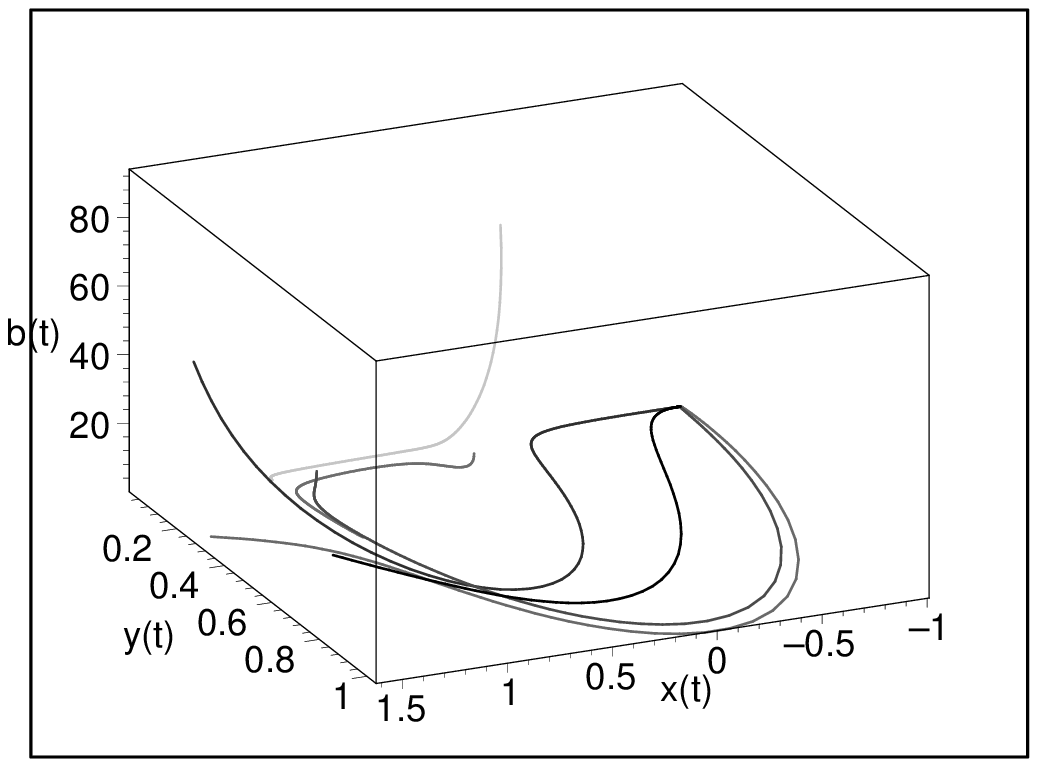}
\includegraphics[width=4.0cm,height=3.5cm]{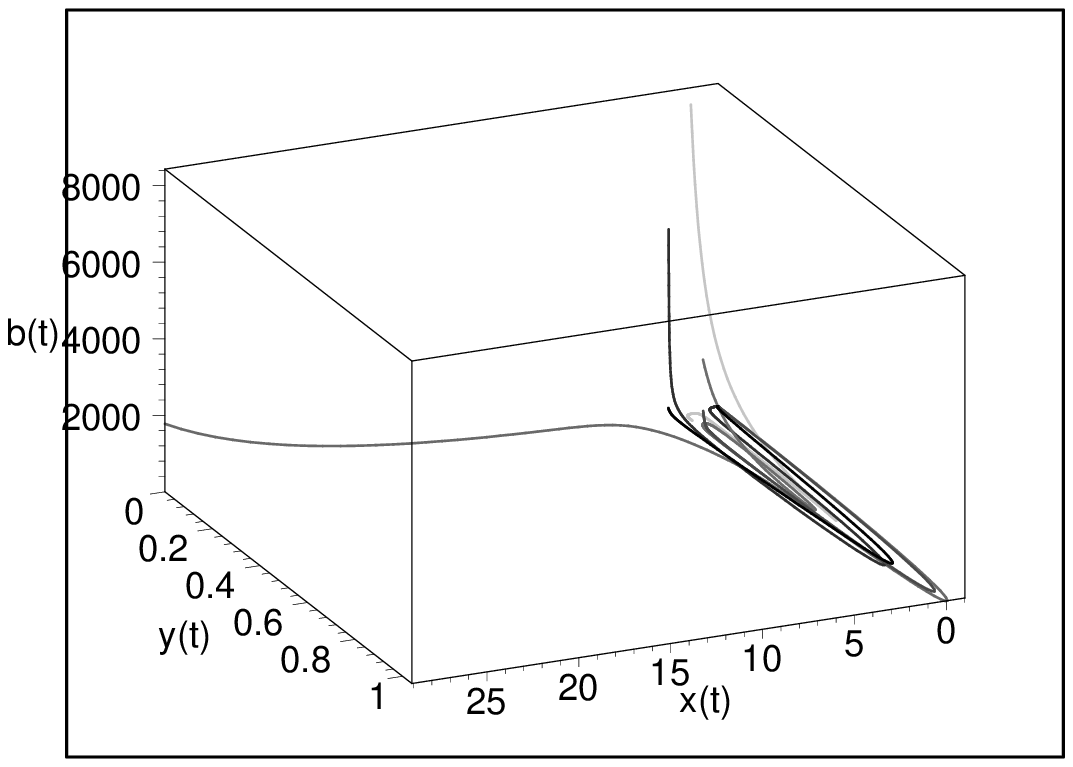}
\includegraphics[width=4.0cm,height=3.5cm]{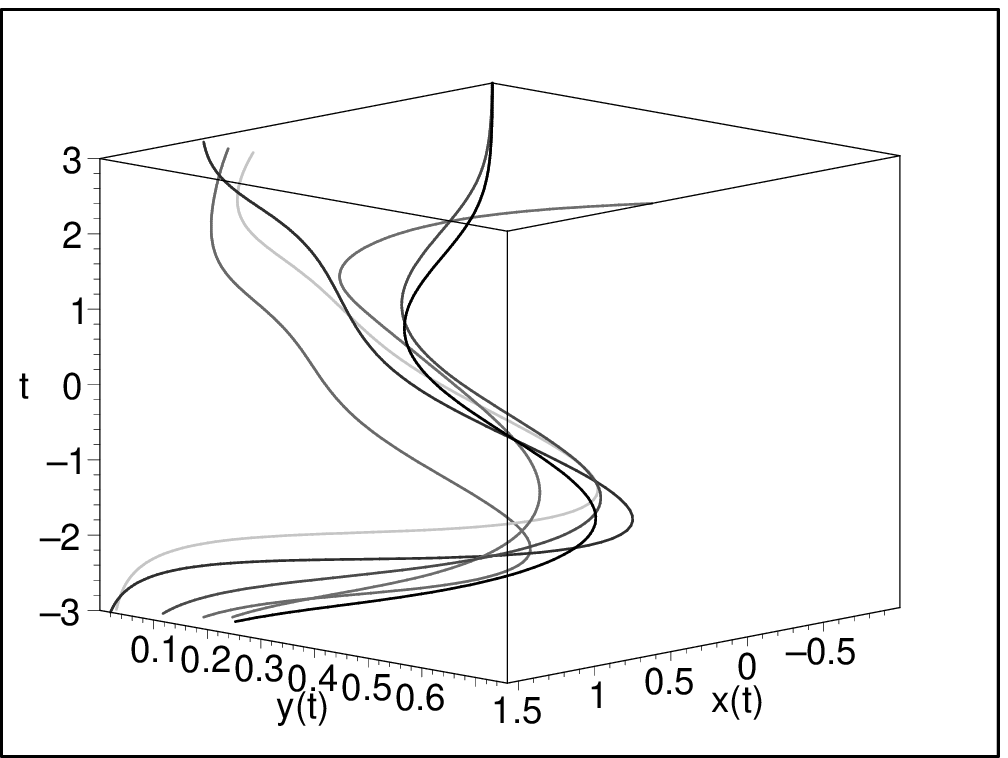}
\includegraphics[width=4.0cm,height=3.5cm]{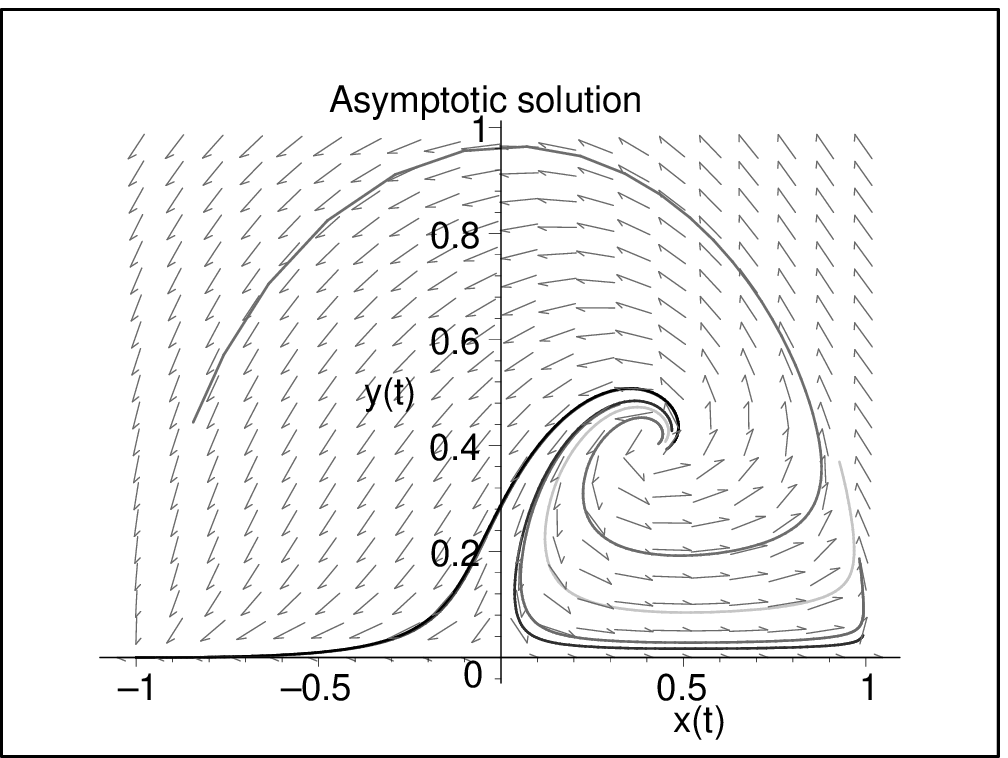}
\vspace{0.3cm}
\caption{Trajectories in phase space for different sets of initial conditions, for $\lambda=3$ in the exponent (upper part of the figure). In the lower part of the figure we show the flux in time (picture to the left), and projection on to the plane $(x,y,0)$ (picture to the right). Due to the ignorance of the brane effects in the projection, the matter-scaling solution seems to be the late-time attractor.}\label{fig2}
\end{centering}
\end{figure}

The critical points of (\ref{22}-\ref{25}) are summarized in the table \ref{tab1}. These coincide with the ones found in reference \cite{wands} as it should be. The point $P_1=(x,y,b)=(0,0,0)$ corresponds to the matter-dominated solution (recall that, in this case $l^2=1$): $H^2=\rho_m/3\mu^2$. The kinetic energy-dominated solutions ($x^2=1,\;l^2=0\;\Rightarrow\;H^2=\dot\phi^2/6\mu^2$) correspond to the points $P_2^\pm$ in Tab.\ref{tab1}. These are always unstable source points in phase space, i. e., the past attractor for every trajectory in the phase space.

The remaining critical points are the scalar field dominated solution (point $P_4$) and the matter-scaling solution $P_3$. In the first case, since $l^2=b^2=0$, then $x^2+y^2=1\;\Rightarrow\;H^2=(\dot\phi^2/2+V)/3\mu^2$. This solution exists whenever $\lambda^2<6$ and is accelerating if $\lambda^2<2$.  The matter-scaling solution (critical point $P_3$) exists for $\lambda^2>3$. In this case $x=y$, $l^2=1-2x^2$. The scalar field fluid mimics dust matter ($\omega_\phi=\omega_m=0$). 

Unlike the standard result in \cite{wands}, both solutions (scalar field-dominated and matter-scaling critical points) represent saddle points in the phase space of the DGP model. Looking at the projections of the phase space onto the plane $(x,y,0)$ -- the plane associated with four-dimensional behavior in the lower figures in Fig. \ref{fig1} and Fig. \ref{fig2} -- so that we are ignoring higher-dimensional effects, one can think that these solutions, depending on the values of the free parameters, can be late-time attractor points in phase space (the standard result in \cite{wands}). However, since in the DGP model higher-dimensional effects modify the late-time dynamics, the actual situation is very different. At late times the trajectories in phase space leave the plane $(x,y,0)$ and asymptotically approach to increasingly large values of the parameter $b$ -- associated with the DGP brane effects -- and of the scalar field kinetic energy $\dot\phi^2/2$, so that there is no isolated critical point in the phase space that could be associated with a unique future (late-time) attractor. The phase space pictures in the figures \ref{fig1} and \ref{fig2}, show this very interesting fact. It is apparent that the phase space trajectories leave the $(x,y,0)$-plane at different places (and times), and these approach to different regions of the 3D phase space. It is clear, also, that the critical points associated with the matter-scaling solution, and with the scalar field-dominated solution, can be only saddle points in the phase space of the DGP cosmological model. 

We can conclude this comment by identifying the source of the incorrect result reported in reference \cite{zhang}, regarding the existence of matter-scaling critical points in a dust universe in the Minkowski cosmological phase of the DGP model with a scalar field trapped on the brane: the inaccurate identification of the phase space corresponding to the model. The existence of matter-scaling solutions is expected from the beginnig since, in the 4-dimensional limit when standard Friedmann behavior is recovered, we are left with the case studied in the reference \cite{wands}, where these solutions were identified as critical points in phase space. That is true even if it is expected that the stability of these points is modified by the infra-red (DGP) brane effects. The kinetic energy-dominated solution is always the past attractor (as in \cite{wands}) since, at early times, the brane effects can be safely ignored so that the standard cosmological dynamics is not modified. 

A graphic illustration of the above discussed features is given in the figures \ref{fig1} and \ref{fig2} above, where the phase space trajectories uncover the main features of the dinamical system of interest in \cite{zhang}.

This work was partly supported by CONACyT M\'exico, under grants 49865-F, 54576-F, 56159-F, 49924-J, 52327, and by grant number I0101/131/07 C-234/07, Instituto Avanzado de Cosmologia (IAC) collaboration. R G-S acknowledges partial support from COFAA-IPN and EDI-IPN grants, and SIP-IPN 20080759. I Q aknowledges also the MES of Cuba for partial support of the research.

\end{document}